\documentstyle[12pt]{article}

\parskip=\medskipamount                 
\begin{document}
\bibliographystyle{unsrt}
\def\be{\begin{equation}}
\def\ee{\end{equation}}
\def\su3c{SU(3)_c}
\def\da{^{\dagger}}
\def\bc{\begin{center}}
\def\ec{\end{center}}
\begin{center}
{\bf COLOR:  FROM BARYON SPECTROSCOPY TO QCD}\\
[.1in]
{\bf O.W. Greenberg\footnote{Supported in
part by the National Science Foundation.
e-mail addresses: greenberg@umdhep (bitnet);
umdhep::greenberg (decnet);
greenberg@umdep1.umd.edu (internet).}}\\
{\it Center for Theoretical Physics\\
 Department of Physics and Astronomy\\
University of Maryland\\
College Park Maryland 20742-4111, U.S.A.}
\end{center}
\vspace{.2in}
\begin{center}
Talk presented at the session ``Thirty years of the quark model''\\
at Baryons '92,  June 1-4, 1992, Yale University.\\
To appear in the Proceedings.\\
University of Maryland Preprint No. 92-241
\end{center}
\vspace{.2in}
\begin{center}
{\bf Abstract}
\end{center}

I review the
discovery of the color degree of freedom in
hadronic physics, and the developments which led from that discovery to the
local gauge theory of color, quantum chromodynamics.

\bc
{\bf 1. DEDICATION}
\ec

I dedicate this talk to the memory of Feza G\"ursey, a person of great kindness
and great insight, whom we miss very much at this conference.  I first met
Feza at a seminar he gave
at Brandeis University in 1957.  A few years later, in 1962, Feza organized
a wonderful Summer School in what was then called the Robert College in Bebek,
near Istanbul.  The participants spent about a month in a lovely
setting overlooking the Bosphorus, eating meals together, attending and giving
lectures, and discussing physics at all hours.  Suha G\"ursey
charmed everybody with her gracious hospitality.  Feza, in addition to
providing his own insights, had assembled a stellar set of
lecturers including Guilio
Racah, Eugene
Wigner, Louis Michel, Sheldon Glashow, Yoichiro Nambu and Abdus Salam.  The
proceedings are still worth reading.\cite{1}

\bc
{\bf 2. THE $SU(6)$ OF G\"URSEY AND RADICATI}
\ec

Since this session is devoted to the history of the quark model,
I will give some personal reminiscences about some of the early work on
the model.  I first heard rumors about the idea of fractionally charged quarks
in the winter of 1963-1964.  One of my first reactions was to wonder why
three quarks or one quark and one
antiquark should be the only combinations which would stick together, a
question not addressed by the original version of the model due to Murray
Gell-Mann\cite{2} and to George Zweig.\cite{3}

     In the summer of 1964, I returned to Maryland from
France where I had collaborated with Albert Messiah on a
series of articles on parastatistics.  Although I was
scheduled to spend the Fall semester of 1964 at the
Institute for Advanced Study and would not be at Maryland
for the Fall semester seminars, my colleagues asked me to invite Feza,
who was then at Brookhaven National Laboratory,
to give a seminar at Maryland.  When I spoke with Feza, he
mentioned that he had just finished some interesting work
with Luigi Radicati, that an article on this work would soon
appear in Physical Review Letters, and that I should read it
when it appeared.  When I reached Princeton, I
found that the paper of G\"ursey and Radicati\cite{4} on
relativistic SU(6) was the center of interest.  Several people at the
Institute started working on developments of the idea of
higher relativistic symmetries,  among them Korkut Bardakci, John
Cornwall, Peter Freund and Benjamin Lee.  I eagerly read the G\"ursey-Radicati
paper.  The main idea was that the flavor degree of freedom of quarks,
$q \sim (u,d,s)$, in $SU(3)_f$ could be combined with their spin degree of
freedom, $(1/2) \sim (\uparrow, \downarrow)$, in $SU(2)_S$ to be a
\begin{equation}
{\bf 6}=(q~ 1/2) \sim (u_{\uparrow}, u_{\downarrow}, d_{\uparrow},
d_{\downarrow},
s_{\uparrow}, s_{\downarrow}) ~{\rm in} ~SU(6)_{fS}.
\end{equation}
This ${\bf 6}$ reduces to the
direct product of a quark triplet and a spin doublet under
\begin{equation}
SU(6)_{fS}\rightarrow SU(3)_f \times SU(2)_S=(u,d,s,)\times (\uparrow,
\downarrow)=({\bf 3}, {\bf 1/2}).
\end{equation}
The G\"ursey-Radicati $SU(6)$
theory successfully places the mesons $\sim q {\bar q}$ in
\begin{equation}
{\bf 6}\otimes {\bf 6}^{\star}={\bf 1}+{\bf 35},
\ee
\be
{\bf 35} \rightarrow ({\bf 8}, {\bf 0})+({\bf 1}+{\bf 8}, {\bf 1})~ {\rm
under}~
SU(6)_{fS} \rightarrow SU(3)_f \times SU(2)_S
\end{equation}
The well-known pseudoscalar meson octet
$(K^+,K^0,\pi^+,\pi^0,\pi^-,\eta^0,{\bar K}^0,{\bar K}^-)$ and the vector meson
nonet $(K^{\star~+},K^{\star~0},\phi^0,\rho^+,\rho^0,\rho^-,\omega^0, {\bar
K}^{\star~0}{\bar K}^{\star~-})$ are unified.  For the baryons $\sim qqq$ in
\begin{equation}
{\bf 6}\otimes {\bf 6}\otimes {\bf 6}={\bf 56}+{\bf 70}+{\bf 20},
\end{equation}
G\"ursey and Radicati choose the ${\bf 56}$ which includes the known nucleon
octet and delta decuplet
\begin{equation}
{\bf 56}\rightarrow ({\bf 8},{\bf 1/2})+({\bf 10},{\bf 3/2})~ {\rm under}~
SU(6)_{fS} \rightarrow SU(3)_f \times SU(2)_S,
\end{equation}
where the nucleon octet is $(p^+,n^0,\Lambda^0,\Sigma^+,\Sigma^0,\Sigma^-,
\Xi^0,\Xi^-)$ and the delta decuplet is\\
$(\Delta^{++},\Delta^+,\Delta^0,\Delta^-,
Y^{\star~+}_1,Y^{\star~0}_1,Y^{\star~-}_1,\Xi^{\star~0},\Xi^{\star~-},
\Omega^-)$.

\bc
{\bf 3. THE STATISTICS PARADOX}
\ec

I noticed that the  representation
which fit  the  low-lying  baryons, the  ${\bf 56}$,  had the  three quarks in
a
symmetric state under permutations, in  contradiction with what one would
expect
from the assumption that quarks have  spin 1/2 which would require they obey
the
Pauli exclusion  principle. The idea of  an $SU(6)_{fS}$
symmetry which combines the
$SU(3)_f$ of  the
then-known  three flavors  with the  $SU(2)_S$ of  spin was found
independently by Bunji Sakita\cite{5} and  by Zweig.\cite{6}
Because I had spent a good
part of the previous two years working on parastatistics,  I immediately
thought
about the possibility of fixing up the  quark statistics by assuming quarks
obey
parafermi statistics of  order three, since then up  to three quarks can be in
a
symmetric state.  I knew that the celebrated spin-statistics theorem, usually
stated ``Particles which obey Bose statistics must have integer spin and
particles which obey Fermi statistics must have odd-half-integer spin''
should be
amended to read ``{\it Given the choice between} parabose and parafermi
statistics, each of integer order $p$,
particles which obey parabose statistics must have integer spin and
particles which obey parafermi statistics must have odd-half-integer
spin.''\cite{7}
The cases for $p=1$ are the usual Bose and Fermi statistics.
The order of the parastatistics for quarks in baryons
is  uniquely fixed by the two
requirements that the  three quarks be in a  symmetric state and that the
proton
and other  baryons not  have any  additional  degeneracy besides  the known
spin
degeneracy.  It was easy  to show that  the  composite state in  which the
three
quarks are  symmetric under  permutations is an  effective  fermion. It was
also
clear that the  statistics  issue did not enter  for mesons, since they are
composed of two dissimilar particles, namely a quark and an antiquark.

I will give only the briefest description of parastatistics,\cite{8}
using the example
of parafermi statistics which is relevant here.  Let $q$ be a parafermi field
of
order $p$.  Then $q$ can be expanded in terms of $p$ ``Green component'' fields
$q^{(\alpha)}$,
\be
q=\sum_{\alpha=1}^{p} q^{(\alpha)},
\ee
\be
[q^{(\alpha)}(x),q^{(\alpha)~\dagger}(y)]_+=\delta({\bf x}-{\bf y}),x^0=y^0,
\ee
\be
[q^{(\alpha)}(x),q^{(\beta)}(y)]_-=0, x^0=y^0,\alpha \neq \beta,
\ee
where $[A,B]_{\pm}=AB \pm BA$.  Thus the Green components behave as fermions
for
the same value of the Green index and as bosons for different values of this
index.  I emphasize that parastatistics is a perfectly valid local,
relativistic
theory with positive probabilities.  In fact, parastatistics is equivalent as a
classification symmetry to a
theory with a hidden $p$-valued degree of freedom,\cite{9,10}
which, for the hadrons, we now call ``color.''
The composite Fermi baryon creation operator
can be written in terms of the quark fields as
\be
B^{\dagger}_{abc}=[[q_a^{\dagger},q_b^{\dagger}]_+,q_c^{\dagger}]_+
=4\sum_{\alpha\neq \beta\neq \gamma\neq \alpha=1}^3 q_a^{(\alpha)\dagger}
q_b^{(\beta)\dagger}q_c^{(\gamma)\dagger},
\ee
where $a,b,c$ are the ``visible'' space-flavor-spin
indices and $\alpha, \beta, \gamma$
are the hidden ``para'' or color indices.  The boson composites in the
parafermi
theory are in one-to-one correspondence with the color-singlet mesons in the
$\su3c$ theory and the fermion composites are in one-to-one correspondence with
the color-singlet baryons.

At least five possible solutions to the paradox of the apparent violation of
the
spin-statistics theorem were considered:\\
1.  Quarks are just a mathematical fiction, not real particles, so the
spin-statistics theorem can be ignored.  Some theorists considered the ``real''
physics to lie in the algebras constructed using certain fields.  By analogy
with haute cuisine, in which hummingbirds' tongues are
cooked between two slices of veal after which the veal is thrown away,
once the algebra was constructed, the fields and any associated objects could
be
discarded.  I unequivocably took the point of view that quarks are real
particles.\\
2.  Quarks are indeed fermions; the statistics paradox is removed by the
ground-state baryons
having a totally antisymmetric space wavefunction.  The simplest polynomial in
the quark coordinates multiplying the exponentially decreasing factor is
$({\bf x}_1^2-{\bf x}_2^2)({\bf x}_2^2-{\bf x}_3^2)({\bf x}_3^2-{\bf x}_1^2)$,
since the state has to have $L=0$.  (The polynomial ${\bf x}_1\times{\bf x}_2
\cdot{\bf x}_3$ vanishes identically because the coordinates are linearly
dependent.)  Such a ground-state wavefunction has two problems: (a) It is not
clear what to choose for excited states.  The suggestion that $q{\bar q}$ pairs
should be added for excited states leads to ``exploding'' $SU(3)_f$
representations which are not observed.  (b) Ashok
Mitra and Rabi Majumdar\cite{11}
and Rodney Kreps and Johann de Swart\cite{12}
pointed out that zeroes in the ground-state wavefunction would lead
to zeroes in the proton form factors, which also are not observed.\\
3.  Quarks obey parastatistics.  I am discussing this possibility here.\\
4.  There are three equivalent triplets which belong to a ${\bf 3}$ of $\su3c$.
The new three-valued degree of freedom can be chosen antisymmetric to resolve
the statistics paradox.
As a new degree of freedom, this is equivalent to parastatistics.\\
5.  There are three different triplets; this degree of freedom can also be
chosen antisymmetric and the charges can be chosen integral.
Moo Young Han and Yoichiro Nambu proposed this model\cite{13} in 1965.
I discuss this possibility below.

\bc
{\bf 4. MAGNETIC MOMENTS OF THE PROTON AND NEUTRON}
\ec

Ben Lee showed me the manuscript of
his article with Mirza B\'eg and Abraham Pais\cite{14}
which derived the ratio of the
magnetic moments of the proton and neutron, the famous
$\mu_p/\mu_n=-3/2$, which is accurate to $3\%$.
Previous work on these magnetic moments had assumed that in lowest
approximation
the proton and neutron are pointlike Dirac particles, in which case this ratio
is infinity (but nobody considered the ratio), and tried to get the
observed magnetic moments by adding anomalous contributions
coming from strong interaction
effects, such as pion clouds.  Not only was this complicated; it also totally
failed to
account for the observed moments.  No one had even noticed that the ratio of
the
total moments was such a simple number.  The B\'eg, Lee, Pais paper used pure
relativistic $SU(6)$ group theory to derive this result.
I was very impressed by
this calculation and became convinced that the quark model
was correct in a very concrete sense:  that quarks are real
and that the hadrons are literally made of them.

I realized  that the  parastatistics scheme is  equivalent for baryons to a
model in which the quarks obey Bose  statistics and could be represented by
Bose
annihilation and  creation operators,  and I set up the  S-wave quark states
for
the proton and  neutron in each spin   state. Using these  states it was easy
to
calculate the  magnetic moments  of the proton  and neutron  and, in
particular,
their ratio. In this way I recovered the B\'eg, Lee, Pais result on the basis
of
a concrete model of quarks.\cite{15}

I remind you of this elementary calculation here.
As I just mentioned above, once you take care of the statistics for the
spin-flavor degrees of freedom using
parafermi quarks of order three, you can use Bose operators for the spin-$1/2$
quarks.  The key observation is that $O_{AB}\equiv
\epsilon^{ab}a^{\dagger}_{Aa}
a^{\dagger}_{Bb}$, where $A,B$ stand for the flavor indices with $1=u,~2=d,~
3=s$ and $a,b$ stand for the spin indices with $1={\uparrow},
{}~2={\downarrow}$,
 has strangeness zero and the isospin of an antisymmetric state
of $q_A$ and $q_B$.  Thus  $O_{AB}$ serves as the
``core'' of all the states in
the nucleon octet
(except $\Sigma^0$, which can be made from $\Sigma^+$ using the
isospin-lowering operator $I_-$), with the third quark carrying the net spin
and
isospin of the baryon.  Then, writing out the terms with
the $\epsilon$, the proton is
\begin{equation}
|p^+_{\uparrow} \rangle={\frac{1}{\sqrt{3}}}u^{\dagger}_{\uparrow}
(u^{\dagger}_{\uparrow}d^{\dagger}_{\downarrow}
-u^{\dagger}_{\downarrow}d^{\dagger}_{\uparrow})|0\rangle   \label{p}
\end{equation}
and the neutron is
\begin{equation}
|n^0_{\uparrow} \rangle={\frac{1}{\sqrt{3}}}d^{\dagger}_{\uparrow}
(u^{\dagger}_{\uparrow}d^{\dagger}_{\downarrow}
-u^{\dagger}_{\downarrow}d^{\dagger}_{\uparrow})|0\rangle.  \label{n}
\end{equation}
The magnetic moment of a baryon $B$ is
\begin{equation}
\mu_B=\langle B_{\uparrow}|\mu_3|B_{\uparrow}\rangle,  \label{mu}
\end{equation}
where
\begin{equation}
\mu_3=2\mu_0\sum_q Q_qS_q,~~ Q_q=(\frac{2}{3},-\frac{1}{3},-\frac{1}{3})
\label{mu3}
\end{equation}
is the charge of the quarks
$(u,d,s)$ and $S_q$ is the $z$-component of the spin of quark $q$.
{}From (\ref{mu}) and (\ref{mu3}), the proton magnetic moment is
\begin{equation}
\mu_p=2\mu_0 \cdot \frac{1}{3} \{2[\frac{2}{3} \cdot 1+(-\frac{1}{3}) \cdot
(-\frac{1}{2})]+[(-\frac{1}{3}) \cdot \frac {1}{2}]\}=\mu_0;
\end{equation}
the two square brackets inside the curly bracket come from the two
non-interfering terms in the proton state, the two multiplying
the first square bracket
is a normalization factor from the two spin up $u$ quarks, the terms inside
the first square bracket reflect the charge and spin of the $u$ and $d$
quarks and
similar comments hold for the second square bracket.  The analogous calculation
for the neutron magnetic moment gives
\begin{equation}
\mu_n=-\frac{2}{3} \mu_0.
\end{equation}
Thus the evaluation of the magnetic moments is just a matter of counting.
The ratio of the moments is $-3/2$, and the calculation of the individual
moments can be
interpreted as just adding up the Dirac moments of the quarks in $S$ states
with the spin-flavor wavefunction given by the ${\bf 56}$,
provided the quarks have constituent masses of $m_N/2.79=336 MeV.$

\bc
{\bf 5. THE SYMMETRIC QUARK MODEL FOR BARYONS}
\ec

I felt that something more had to be done to justify the idea that quarks obey
parastatistics, and decided that I should study the orbitally-excited
baryon states which would
lie just above the ground-state ${\bf 56}$.  At this stage, I felt the need of
a
collaborator and approached Ben Lee, asking if he would join me in this work.
After a day or so, Ben said that he was too busy working on relativistic
$SU(6)$
inspired by the paper of G\"ursey and Radicati with some other people at the
Institute.  I then decided to pursue this work by myself.
My development of the paraquark model benefited from discussions with
Alex Dragt, Peter Freund, Ben Lee and Samuel MacDowell.  I would also like to
acknowledge the attentive ear of Guido Sandri.  The Institute is an ideal
place for such work, free from distractions,
yet with many stimulating people to talk to.
I worked very intensively at this project.
In three weeks, I wrote and submitted my paper to Physical Review Letters.
In the evening of the day
I had put the manuscript in the mail, I realized that I had ignored
the question of center-of-mass motion, so that some of the excited states I had
predicted were either wholly or in part just the ground state ${\bf 56}$ in
motion.  For example, if $a^{\dagger}$'s create $S$-wave quarks and
$b^{\dagger}$'s create $P$-wave quarks, $(a\da a\da b\da +a\da b\da a\da+
b\da a\da a\da)|0\rangle$ is just the ground-state ${\bf 56}$;
while $a\da (a\da b\da-b\da a\da)|0\rangle$ and
$(2b\da a\da a\da -a\da a\da b\da
-a\da b\da a\da)|0\rangle$ are the true first excited states, the $({\bf 70},
L^P=1^-)$.  I called Sydney Meshkov at the
National Bureau of Standards to get some
clues as to how to eliminate these spurious states, and spent the remainder of
the night rewriting the paper.  I submitted the revised paper the next day.  In
the end it turned out that some of my states still had some partial admixture
of
center-of-mass motion.  The correct states found later by Gabriel Karl and
Edward Obryk\cite{16}  use the
traceless modes, rather than the single-particle states.
Possibly Gabriel will say something about that later this evening.  In those
days, Samuel
Goudsmit was editor of Physical Review Letters, and he made decisions
about publication without protracted debates among referees, divisional
associate editors and authors.  In three more weeks, my paper was in print.
\cite{15}

Naturally, I gave a copy of my paper to Robert
Oppenheimer, who was Director of the
Institute.  A week or so later, there was an Eastern Theoretical Physics
Conference at the new Center of Adult Education at my home institution, the
University of Maryland.  I recall verbatim his comments when I encountered him
at the conference.  He said ``Your paper is beautiful,'' and I went into an
excited state; then he continued
``but I don't believe a word of it.'' That brought me back to my ground state.

Later that fall, I was asked to give a seminar at Harvard.  After the seminar,
in the parking lot coming back from lunch,
Julian Schwinger remarked that the extra degree of freedom implicit in the
parastatistics model ought to play a dynamical role.  A prescient comment,
which
to my chagrin I did not follow up.

\bc
{\bf 6. THE INTRODUCTION OF COLOR $SU(3)$}
\ec

Han and Nambu\cite{13} were the first who explicitly introduced the color
$SU(3)_c$ symmetry
which is implicit in the parastatistics model.  One of their motivations was to
avoid fractional quark charges, so they arranged for their three flavor
triplets
to have different electric charges : $(1,0,0); (1,0,0); (0, -1,-1)$  so that
they are distinguishable.  Averaging
over the charges for each flavor gives the fractional charges of the original
quark model.  They proposed that the forces between
these quarks would be mediated by the
exchange of an $\su3c$-octet of gauge vector mesons, thus giving the hidden
color degree of freedom a dynamical role, and showed that such forces would
make the $\su3c$-singlets be the ground-state particles, to be identified with
the known baryons and mesons.

QCD consists of two
statements:  (a) there is a hidden three-valued degree of freedom carried by
quarks and (b) this degree of freedom is associated with a local $\su3c$
gauge theory with its octet of vector mesons serving to
mediate the strong force between the quarks.
The parastatistics model, translated to explicit triplets, requires identical
triplets and accounts for (a).  The Han-Nambu model accounts for (b).
Thus the union of the correct parts of these two models
is the basis of QCD.

While preparing for this session, I looked into the origin of the
use of the word ``color'' for the gauged $SU(3)_c$ degree of freedom.
Bram Pais,\cite{17} in
a discussion in
the Erice summer school of 1965, was
the first to use color in this way.
Donald Lichtenberg\cite{18} also used color with
this meaning in his book published in 1970.
Color in this sense came into
general use following the articles by Gell-Mann and by William
Bardeen, Harald Fritzsch and Gell-Mann.\cite{19}

In 1965, Daniel Zwanziger and I
followed up the work of Han and Nambu\cite{13} to account
for the fact that only the combinations of quarks and antiquarks, $qqq$ and $q
\bar{q}$, which form baryons and mesons appear in Nature.
Even now there are
no definitely established states beyond these.  We surveyed the existing models
and tried to construct new models to account for this fact.  We noticed that
the
parastatistics model is equivalent to the three-triplet or color model as far
as
the spectrum of color-singlet states is concerned.  As mentioned above,
the composite
particles which are bosons in the parastatistics model are color-singlet mesons
in the color model and the particles which are fermions in the parastatistics
model are color-singlet baryons in the color model.
We found convincing arguments
that only the parastatistics model and its equivalent,
the color model, can account
for the ``saturation'' found in Nature.\cite{9}

\bc
{\bf 7. DEVELOPMENT OF THE\\ SYMMETRIC QUARK MODEL FOR BARYONS}
\ec

In 1966 I developed the idea, which I had sketched in my 1964 paper,
that the forces between quarks should mainly be
two-body forces and applied this simplifying assumption to the baryons in the
ground-state $({\bf 56, L^P=0^+})$
and the first excited supermultiplet, the $({\bf 70,
L^P=1^-}$),
which I had earlier noted should have odd parity.
I asked one of our postdoctoral fellows
at Maryland, Marvin Resnikoff, to join me on this project.

I will briefly describe the simplifications which follow from the two-body
force
assumption.  In $SU(6)_{fS}$, the quark is a ${\bf 6}$ which reduces to a
$({\bf 3}, {\bf 2})$ under
$SU(6)_{fS} \rightarrow SU(3)_f \times SU(2)_S$, where
the $SU(3)_f$ is the flavor symmetry of the $u,d$ and $s$ quarks and the
$SU(2)_S$ is the spin symmetry of the quarks.  As already mentioned,
the quarks in a baryon must be
totally symmetric under permutations of all their degrees of freedom except
color.  For the baryon ${\bf 56}$,
in which the space wave function is symmetric,
the two-body forces act on the symmetric part of ${\bf 6} \times {\bf 6}$,
which is the ${\bf 21}$ of $SU(6)$.  Thus the two-body forces must be in
${\bf 21} \times {\bf 21^{\star}}$, which reduces to ${\bf 1}+{\bf 35}+
{\bf 405}$.  The one-body forces or masses must be in ${\bf 1}+{\bf 35}$.
Further, the forces must be singlets under spin and isospin and
have hypercharge zero.  Finally, one assumes dominance of operators in a flavor
octet.  With these conditions, there are precisely four possible mass
operators,
and they lead to the G\"ursey-Radicati mass formula,
\be
M=M_0+M_1Y+M_2C_2^{(3)}+M_3[I(I+1)-\frac{1}{4}Y^2],
\ee
where $I, Y~{\rm and}~ C_2^{(3)}$ are the hypercharge, isospin and
quadratic Casimir of $SU(3)_f$,
respectively.  This gives four mass relations,
\be
N+\Xi=\frac{1}{2}(3\Lambda+\Sigma),
{}~\Omega-\Xi^{\star}=\Xi^{\star}-Y_1^{\star}=Y_1^{\star}-\Delta=\Xi-\Sigma
\ee
among the eight masses in the $({\bf 56, 0}^+)$.
The last of these, which relates mass
splittings in the octet and decuplet, is the most interesting.
By contrast, without two-body dominance the most general mass
operator acting on the ${\bf 56}$ is in ${\bf 56} \times {\bf 56^{\star}}=
{\bf 1}+{\bf 35}+{\bf 405}+{\bf 2695}$.  Since there are eight admissible
mass operators in these representations, no mass relations follow.  Resnikoff
and I applied these ideas to the $({\bf 56, 0^+})$ and the $({\bf 70, 1^-})$,
fitting all the available baryon resonances and predicting the
remaining ones in these representations.  We felt that we should study the
resonances with $S$- and $P$-wave quarks
in a unified way.  While we were completing
our work, Pedro Federman, Hector Rubinstein and Igal Talmi\cite{20}
published a paper on the baryons in the $({\bf 56, 0^+})$
from a similar point of view.  Our fit to the resonances convinced me,
but at the time not too many others, that the ``symmetric'' quark model is
correct for baryons, and therefore that color or its equivalent, parastatistics
is present in Nature.  I presented our results at the Washington
meeting of the American Physical Society in April of 1966.  The last step in
the
calculation was to get the best fit to all the resonances.  Since there are
resonances which can be superpositions of three states in the model with the
same values of $I, Y~ {\rm and}~ S$, I used
a primitive Basic program to do the
fitting.  I recall running the program on a computer at Dartmouth over
telephone
lines which would fail from time to time.  I sent an expository
article based on
my talk reviewing the state of hadron spectroscopy to Physics Today, which
aged the manuscript for about a year and then rejected it.
Resnikoff and I published a technical paper on our work in Physical
Review.\cite{21} A
year or so later, I supervised a thesis of Dattaprasad
Divgi which extended the results on baryon spectroscopy and also calculated
baryon decays.\cite{22}
Richard Dalitz and his students extended the study
of baryon spectroscopy to higher supermultiplets.\cite{23}
With the advent of QCD, the general group
theory plus two-body force analysis could be replaced by a definite
Hamiltonian for the constituent quarks,
and results from the quantum mechanics of
the Coulomb potential could be applied to baryon spectra.  The pioneers in this
subject were Alvaro De Rujula, Howard Georgi and Sheldon Glashow,\cite{24}
and the people
who developed this work to a fine art were Nathan Isgur and Gabriel
Karl.\cite{25}  I
look forward to Gabriel's talk later this evening.

\bc
{\bf 8. GRADUAL ACCEPTANCE OF THE\\ QUARK MODEL WITH COLOR}
\ec

Although the baryon spectroscopy, the magnetic moment calculation, the quark
counting relations which relate meson-baryon total cross sections to
baryon-baryon total cross sections,\cite{26} the Zweig
rule\cite{3,27}
forbidding decays such as $\phi \rightarrow \rho \pi$ in which no
quarks are
in common in the initial and final states, which were found in the period
1964-1966,
convinced many, including myself, that quarks exist as concrete objects and are
fundamental for hadron physics, others held back.  Some who accepted quarks
held
out against color.  The first rapporteur talk on
hadron spectroscopy which advocated the symmetric quark model for baryons was
given by Haim Harari\cite{28}
at the Vienna conference in 1968.  Some of the holdouts
were convinced by the
success of Feynman's parton model,\cite{29}
with quarks and antiquarks as the partons as suggested by James Bjorken and
Emmanuel Paschos\cite{30}, after
the SLAC deep inelastic scattering experiments in 1969.  We will hear more
about
this from Finn Ravndal.  The understanding that the
$J$/ $\Psi$ and related resonances are charm-anticharm composites in 1974
finally convinced the remaining holdouts.

Several of the effects due to color depend only on color as a classification
symmetry; these, of course, follow equally from the parastatistics model.
Among these are (a) the $\pi^0 \rightarrow \gamma \gamma$ decay rate,\cite{31}
which is proportional to
the square of the number of colors, since the $\pi^0 \rightarrow \gamma \gamma$
amplitude is proportional to the number of quark fields which circulate in the
triangle in the graph for this amplitude and it doesn't matter whether these
quark fields are Green components of an order-3
parafermi field or the components of a
color triplet and (b)
the ratio of the cross section for
electron-positron annihilation to hadrons to the cross section for annihilation
to muon pairs,\cite{32}
which is proportional to the number of colors, because it doesn't
matter whether Green or color components circulate in the loop for the total
hadron cross section.
By contrast, the
property of asymptotic freedom,\cite{33}
that the interaction between quarks falls to zero
at short distance or, equivalently at high energy, requires the gauged theory
of
color $SU(3)_c$, QCD.  We believe that QCD also
leads to permanent confinement of quarks, antiquarks and other color-carrying
objects into colorless $SU(3)_c$-singlets and thus explains the absence of free
quarks with fractional electric charge.  (Parenthetically, later
Kenneth Macrae and I
were able to construct a version of the parastatistics theory which can be
gauged and is equivalent to QCD.\cite{34})

\bc
{\bf 9. GOALS FOR THE FUTURE}
\ec

I conclude by emphasizing that a full understanding of hadrons based directly
on the Lagrangian of QCD is still lacking.  I believe that it is possible to
solve QCD with sufficient accuracy to derive the constituent quark model,
including masses, decays, magnetic moments and other static properties of
hadrons, to demonstrate permanent confinement of quarks and other
color-carrying
objects, to derive chiral symmetry and its breaking in their respective
regimes,
and at the same time to derive the parton model for both deep inelastic
lepton-hadron and high momentum-transfer hadron-hadron scattering.  To realize
these goals using continuum quantum field theory methods
remains a challenge for the future.

\bc
{\bf ACKNOWLEDGEMENTS}
\ec

I thank George Snow and Joseph Sucher for reading the manuscript and for
helpful comments on it.

\bibliographystyle{unsrt}

\begin{thebibliography}{99}
\bibitem{1} {\em Group Theoretical Concepts and Methods in Elementary Particle
Physics}, ed. F. G\"ursey (Gordon and Breach, New York, 1965).  The school is
described in O.W. Greenberg and E.P. Wigner, {\em Physics Today} {\bf 16}
(1963) 62.
\bibitem{2} M. Gell-Mann, {\em Phys. Lett.} {\bf 8} (1964) 214.
\bibitem{3} G. Zweig, CERN Reports 8182/TH.401 and 8419/TH.412 (1964).  The
latter is reprinted in {\em Developments in the Quark Theory of Hadrons},
ed. D.B. Lichtenberg and S.P. Rosen (Hadronic Press, Nonamtum, Mass., 1980).
\bibitem{4} F. G\"ursey and L.A. Radicati, {\em Phys. Rev. Lett.} {\bf 13}
(1964) 173.
\bibitem{5} B. Sakita, {\em Phys. Rev.} {\bf 136B} (1964) 1756.
\bibitem{6} G. Zweig, in {\em Symmetries in Elementary Particle Physics}, ed.
A. Zichichi (Academic, New York, 1965) p.~192.
\bibitem{7} G.F. Dell'Antonio, O.W. Greenberg and E.C.G. Sudarshan, in
{\em Group Theoretical Concepts and Methods in Elementary Particle
Physics}, ed. F. G\"ursey (Gordon and Breach, New York, 1965) p.~403.
\bibitem{8} H.S. Green, {\em Phys. Rev.} {\bf 90} (1953) 270; O.W. Greenberg
and
A.M.L. Messiah, {\em Phys. Rev.} {\bf 138} (1965) B1165.
\bibitem{9} O.W. Greenberg and D. Zwanziger, {\em Phys. Rev.} {\bf 150} (1966)
1177.
\bibitem{10} K. Dr\"uhl, R. Haag and J.E. Roberts, {\em Commun. Math. Phys.}
{\bf 18} (1970) 204.
\bibitem{11} A.N. Mitra and R. Majumdar, {\em Phys. Rev.} {\bf 150} (1966)
1194.
\bibitem{12} R.E. Kreps and J.J. de Swart, {\em Phys. Rev.} {\bf 162} (1967)
1729.
\bibitem{13} Y. Nambu, in {\em Preludes in Theoretical Physics}, ed. A. de
Shalit, H. Feshbach and L. Van Hove (North Holland, Amsterdam, 1966), p.~133;
M.Y. Han and Y. Nambu, {\em Phys. Rev.} {\bf 139} (1965) B1006.  A.
Tavkhelidze,
in {\em High-Energy Physics and Elementary Particles} (Int. Atomic Energy
Agency, Vienna, 1965), also suggested three different triplets
with integral charges, but
did not suggest the gauge interaction.
\bibitem{14} M.A.B. B\'eg, B.W. Lee and A. Pais, {\em Phys. Rev. Lett.} {\bf
13}
(1964) 514, erratum 650.  The magnetic moment ratio was found independently by
B. Sakita, {\em Phys. Rev. Lett.} {\bf 13} (1964) 643.
\bibitem{15} O.W. Greenberg, {\em Phys. Rev. Lett.} {\bf 13} (1964) 598.
\bibitem{16} G. Karl and E. Obryk, {\em Nucl. Phys.} {\bf B8} (1968) 609.
\bibitem{17} A. Pais, in {\em Recent Developments in Particle Symmetries},
ed. A. Zichichi (Academic, New York, 1966) p.~406.
\bibitem{18} D.B. Lichtenberg, {\em Unitary Symmetry and Elementary Particles}
(Academic, New York, 1970) p.~227.
\bibitem{19} M. Gell-Mann, {\em Elementary Particle Physics}, ed. P. Urban
(Springer, Vienna, 1972); {\em Acta Phys. Austriaca Supp.} {\bf 9} (1972) 733;
W.A. Bardeen, H. Fritzsch and M. Gell-Mann, in {\em Scale and Conformal
Symmetry in Hadron Physics}, ed. R. Gatto (Wiley, New York, 1973) p~139.
\bibitem{20} P. Federman, H.R. Rubinstein and I. Talmi, {\em Phys. Lett.} {\bf
22} (1966) 203.
\bibitem{21} O.W. Greenberg and M. Resnikoff, {\em Phys. Rev.} {\bf 163}
(1967) 1844.
\bibitem{22} D.R. Divgi, {\em University of Maryland Thesis} (1968) and
{\em Phys. Rev.} {\bf 175} (1968) 2027;
D.R. Divgi and O.W. Greenberg, {\em Phys.
Rev.} {\bf 175} (1968) 2024, erratum {\bf 178} (1969) 2487.
\bibitem{23} R.H. Dalitz, in {\em Fundamentals of Quark Models},
ed. I. M Barbour, A.T. Davies(Scottish
Univ. Summer School, Edingburgh, 1977).
\bibitem{24} A. DeR\'ujula, H. Georgi and S.L. Glashow {\em Phys. Rev. D}
{\bf 12} (1975) 147.
\bibitem{25} N. Isgur and G. Karl, in {\em Recent Developments in High-Energy
Physics}, ed. A. Perlmutter and L.F. Scott (Plenum, New York, 1980) p.~61.
\bibitem{26} E.M. Levin and L.L. Frankfurt, {\em JETP Lett.} {\bf 2} (1965) 199
(English translation).
\bibitem{27} S. Okubo, {\em Phys. Lett.} {\bf 5} (1963) 165.
\bibitem{28} H. Harari, in {\em 14th Int. Conf. on High-Energy Physics},
ed. J. Prentki and J. Steinberger (CERN,
Geneva, 1968) p.~195.
\bibitem{29} R.P. Feynman, {\em Phys. Rev. Lett.} {\bf 23} (1969) 1415 and
{\em Photon-Hadron Interactions} (W.A. Benjamin, Reading, Mass., 1972).
\bibitem{30} J.D. Bjorken and E.A. Paschos, {\em Phys. Rev.}
{\bf 185} (1969) 1975.
\bibitem{31} S.L. Adler, {\em Phys. Rev.} {\bf 177} (1969) 2426; J. Bell and
R. Jackiw, {\em Nuovo Cimento} {\bf 60A} (1969) 47.
\bibitem{32} N. Cabibbo, G. Parisi and M. Testa, {\em Nuovo Cimento Lett.}
{\bf 4} (1970) 35.
\bibitem{33} D. Gross and F. Wilczek, {\em Phys. Rev. Lett.} {\bf 30} (1973)
1343; H.D. Politzer, ibid 1346.
\bibitem{34} O.W. Greenberg and K.I. Macrae, {\em Nucl. Phys. B} {\bf 219}
(1983) 358.
\end{thebibliography}

\end{document}